# SPAD: Specialized Prefill and Decode Hardware for Disaggregated LLM Inference

Hengrui Zhang
hengrui.zhang@princeton.edu
Princeton University

Pratyush Patel
patelp1@cs.washington.edu
University of Washington

August Ning
aning@princeton.edu
Princeton University

David Wentzlaff
wentzlaf@princeton.edu
Princeton University

## Abstract

Large Language Models (LLMs) have gained popularity in recent years, driving up the demand for inference. LLM inference is composed of two phases with distinct characteristics: a compute-bound prefill phase followed by a memory-bound decode phase. To efficiently serve LLMs, prior work proposes prefill-decode disaggregation to run each phase on separate hardware. However, existing hardware poorly matches the different requirements of each phase. Current datacenter GPUs and TPUs follow a *more-is-better* design philosophy that maximizes compute and memory resources, causing memory bandwidth underutilization in the prefill phase and compute underutilization in the decode phase. Such underutilization directly translates into increased serving costs.

This paper proposes SPAD (Specialized Prefill and Decode hardware), adopting a *less-is-more* methodology to design specialized chips tailored to the distinct characteristics of prefill and decode phases. The proposed Prefill Chips have larger systolic arrays and use cost-effective GDDR memory, whereas the proposed Decode Chips retain high memory bandwidth but reduce compute capacity. Compared to modeled H100s, simulations show that the proposed Prefill Chips deliver 8% higher prefill performance on average at 52% lower hardware cost, while the proposed Decode Chips achieve 97% of the decode performance with 28% lower TDP.

End-to-end simulations on production traces show that SPAD reduces hardware cost by 19%-41% and TDP by 2%-17% compared to modeled baseline clusters while offering the same performance. Even when models and workloads change, SPAD can reallocate either type of chip to run either phase and still achieve 11%-43% lower hardware costs, demonstrating the longevity of the SPAD design.

## 1 Introduction

Large Language Models (LLMs) have become immensely popular due to their advanced capabilities and are being widely adopted in various applications spanning chatbots and code generation tools. However, serving LLMs incurs significant hardware costs. In early 2023, only three months after the introduction of GPT-3.5, it was estimated that ChatGPT cost nearly $700,000 per day to serve with around 30,000 NVIDIA

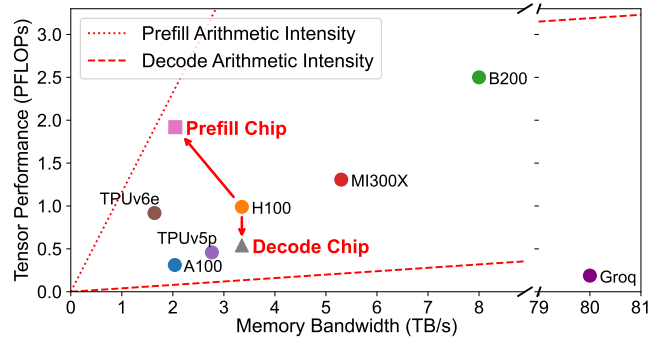

**Figure 1.** Comparison of LLM Serving Hardware. We estimate prefill and decode arithmetic intensities for BLOOM-176B (FP16, sequence length 1024) with batch sizes of 2 and 64 respectively (shown as the dashed lines). Only the FLOPs and memory accesses related to matrix multiplications are included. The tensor performance and memory bandwidth numbers are theoretical values reported by their specifications [8, 15, 16, 26, 42–44].

A100 GPUs [49]. The demand for LLMs has increased rapidly since then. In March 2025, NVIDIA announced that they had received orders for 3.6 million of their newest flagship Blackwell GPUs from cloud providers [38], of which a significant fraction will likely be used to serve LLMs.

Serving LLMs is expensive due to tight latency constraints coupled with high hardware requirements. LLM inference executes in two phases with different computational properties. In the compute-bound *prefill* phase, all tokens in the input prompt are processed in parallel to generate the KV cache and the first output token. In the memory-bound *decode* phase, subsequent output tokens are generated sequentially, where each new token depends on the KV cache state of all previous tokens. This dual-phase nature poses a challenge for existing hardware since each phase effectively utilizes only a subset of the hardware resources.

To improve efficiency, prior work has proposed two broad scheduling techniques. **Co-location**-based schedulings batch the prefill and decode phases of different requests together to improve hardware utilization, leveraging the fact that both phases share the same model weights [4, 69]. However, this approach incurs large tail latencies due to prefill-decode interference [74], leading to a less responsive user experience. **Disaggregation**-based scheduling separates the execution of the prefill and decode phases onto different hardware by using interconnects to transfer the KV cache [51, 74]. Although this approach incurs transfer overheads, it improves overall performance by enabling phase-specific resource management and using hardware that better matches the computational characteristics of each phase.

**Despite these software-level optimizations, the hardware efficiency of serving LLM inference is still fundamentally limited by the mismatch between the workload requirements (i.e., TFLOPS and memory accesses) and hardware resources (i.e., compute capacity and memory bandwidth).** Current datacenter GPU/TPU design philosophy tends to fit as much compute capacity and cache as possible onto a reticle-sized die and pair it with High Bandwidth Memory (HBM) using the advanced CoWoS (Chip-on-Wafer-on-Substrate) packaging technology [28]. The resulting enormous TFLOPs and memory bandwidths make them the most popular hardware platform for serving LLMs. However, this *more-is-better* design philosophy drives up costs for disaggregation-based LLM inference: the high arithmetic intensity of the prefill phase leaves the expensive HBM underutilized, and the low arithmetic intensity of the decode phase leaves the compute capacity underutilized. Our simulations show that reducing the memory bandwidth of a modeled NVIDIA H100 by 40% would only increase prefill latency by 17%. Likewise, the simulated decode latency only increases by 22% if we reduce the compute capacity by half.

**We propose SPAD (Specialized Prefill and Decode hardware), tailoring specialized hardware to the distinct characteristics of prefill and decode phases to improve disaggregation-based LLM serving efficiency.** In contrast to the *more-is-better* design philosophy of GPUs, SPAD embraces a *less-is-more* design philosophy that right-sizes hardware for each phase, while still retaining the ability to run the other phase. For the compute-heavy prefill phase, we propose a specialized Prefill Chip with larger systolic arrays and a cost-effective GDDR-based memory system. For the low-arithmetic-intensity decode phase, we propose an area- and TDP-efficient Decode Chip with smaller systolic arrays and caches. Simulations with LLMCompass [71] show that compared to modeled H100s, our proposed Prefill Chips deliver 8% higher prefill performance on average at 52% lower hardware cost, while our proposed Decode Chips achieve 97% of the decode performance with 28% lower TDP.

We evaluate SPAD by provisioning cost-optimized heterogeneous clusters. End-to-end simulations on production traces for chatbot and code generation applications show that SPAD clusters reduce hardware cost by 19%-41% and TDP by 2%-17% compared to modeled baseline clusters while maintaining the same performance. Even as the models and workloads change, SPAD can reallocate either type of chip to run either phase and still achieve an 11%-43% lower hardware cost, demonstrating the longevity of our design.

In summary, our contributions are as follows:

- Identify the inherent hardware inefficiency of modern GPUs for disaggregated LLM serving. (Section 3)
- Propose SPAD, a heterogeneous system that adopts a *less-is-more* philosophy to design specialized Prefill and Decode Chips to efficiently serve the corresponding phases of LLM inference. (Sections 4 and 5)
- Conduct extensive end-to-end cluster-level simulations demonstrating the cost-effectiveness and longevity of SPAD under various workloads and model architectures. (Sections 6 and 7)

## 2 Background and Related Work

We start by providing an overview of LLM architecture, hardware choices, and software-based serving techniques.

### 2.1 Generative LLMs

**Transformers.** Most modern LLMs like GPT-4 [47], DeepSeek-V3 [20], Llama-3 [25], and Grok-3 [66] are based on the decoder-only transformer architecture [62]. Each transformer block consists of two key components: a self-attention mechanism and a feed-forward neural network. Self-attention enables each token to directly compute relationships with all prior tokens in the sequence. Feed-forward networks (FFNs) process the attention-weighted tensors through linear and non-linear transformations. To support larger model sizes with cheaper inference, a sparse Mixture-of-Experts (MoE) architecture [20] has been adopted, which uses multiple FFNs, called experts, of which only a subset is dynamically activated through a communication-intensive routing mechanism [14].

**Inference Phases.** Generative LLMs operate in two distinct computational phases with different resource requirements. The prefill phase processes the input prompts provided by the user in one forward pass to generate the first output token and the key-value (KV) cache, which facilitates further token generation. Prefill computation is parallelized across all input tokens and has high compute utilization. The decode phase generates subsequent tokens one at a time by running forward passes with the previously generated token along with the KV cache of all prior tokens. This phase is memory-bound because generating each new token requires loading the entire model weights along with the growing KV cache.

**Performance SLOs.** LLM serving usually has tight latency requirements expressed as Service-Level Objectives (SLO). From a user perspective, the prefill *time to first token* (TTFT) measures the latency to receive an initial response, while the *time between tokens* (TBT) measures how quickly the decode phase generates the rest of the response. Both TTFT and TBT are important to ensure an interactive user experience.

## 2.2 Hardware for LLMs

Today, LLMs are mainly served with GPUs and TPUs. To meet high resource demands, such hardware tends to maximize memory and compute capacities as much as possible.

**Memory.** LLM inference has high memory bandwidth and capacity requirements due to the large model sizes and KV cache sizes involved. High-end GPUs and TPUs incorporate high-bandwidth memory (HBM) [8, 15, 44] to meet these needs, and Groq uses SRAMs for even better performance [26], despite needing a large number of chips to meet capacity requirements. LPDDR and CXL memory have also been explored to reduce cost and power usage [48, 71].

**Compute.** Tensor operations like matrix multiplications dominate LLM execution, which has led hardware to adopt specialized components to accelerate their computation. For example, NVIDIA H100s incorporate 528 Tensor Cores to deliver nearly 1000 TFLOPs of FP16 dense matrix multiplication performance [44] and Google TPUs use large systolic arrays specifically designed for matrix computations [30–32, 40]. Non-tensor operations, such as activation or normalization functions, are usually mapped to a more general-purpose SIMD (Single Instruction Multiple Data) or vector units with lower peak performance. These tensor units and vector units can take a significant amount of die area, driving up the hardware manufacturing cost and TDP.

## 2.3 Efficient Serving Techniques

The computational differences between prefill and decode phases cause efficiency issues, inspiring software solutions.

**Co-location.** Traditional serving systems like Orca [69] run requests end-to-end on the same hardware [9, 69], batching them at request or iteration granularity. Recent systems like Sarathi [4], POD-Attention [33], and Nanoflow [75] chunk prefill phases to match hardware compute capacity and batch them with decode phases of different requests to better utilize memory bandwidth. This approach improves hardware utilization and can support very high throughput, but it incurs resource contention between prefill and decode computations [74], causing large tail TTFT and TBT latencies that can violate SLOs.

**Disaggregation.** Splitwise [51] and DistServe [74] disaggregate inference phases across different hardware clusters and use fast interconnects like Infiniband or NVLink to efficiently transfer KV caches between them. This approach eliminates cross-phase interference and enables phase-specific resource management and hardware choices. Prefills can optimize TTFTs by matching hardware compute capabilities, while decodes can optimize throughput by batching more requests

Table 1. LLM Serving Cluster Design Space

| | Spec. | Hetero. | Disagg. | Latency | Throughput | Cost |
|---|---|---|---|---|---|---|
| **Orca [69]** | ✗ | ✗ | ✗ | Variable | Low | High |
| **Sarathi [4]** | ✗ | ✗ | ✗ | Variable | Very High | High |
| **Groq [26]** | ✔ | ✗ | ? | Very Low | ? | ? |
| **DistServe [74]** | ✗ | ✗ | ✔ | Low | High | Med |
| **Splitwise [51]** | ✗ | ✔ | ✔ | Low | High | Med |
| **SPAD** | ✔ | ✔ | ✔ | Low | High | Low |

**Spec**: specialized chip, **Hetero**: heterogeneous, **Disagg**: disaggregation-based scheduling, **Cost**: Cost per goodput. ?: Groq likely runs with small batches due to low memory capacity and has lower throughput.

to improve performance under SLOs. Due to its effectiveness, this idea has been adopted in production systems like NVIDIA Dynamo [45], Mooncake [53], and DeepSeek [20] alongside other optimizations. Some disaggregated systems also adopt co-location-based techniques, such as chunked prefills to better match hardware capacity [45, 73] and mixed batching to handle workload changes and provide better hardware utilization [51].

**Other Techniques.** Recent works improve serving efficiency using a variety of software-based techniques including efficient scheduling [52, 55, 58], memory management [24, 34, 70], kernel optimizations [18, 65, 68], quantization [22, 27, 72], power management [50], etc. These techniques are orthogonal to our work, so we do not discuss them here.

## 2.4 Cluster Designs and Trade-offs

LLM serving clusters can be characterized across three dimensions: hardware specialization, hardware homogeneity, and scheduling. Cluster operators usually choose the design based on hardware availability and workload requirements.

Prior work has explored different points within this design space, as shown in Table 1. Sarathi uses general-purpose GPUs, homogeneous chips, and co-location-based scheduling to enable high utilization with lower management complexity [4]. Clouds built using Google TPUs and Groq leverage specialized hardware to reduce latency and TCO, but probably use the same chips for the entire execution [2, 30]. DistServe disaggregates inference phases on homogeneous GPUs to improve throughput under SLOs [74]. Splitwise further shows that phase-specific hardware, such as H100s for prefills and A100s for decodes, can reduce overall TCO and power consumption [51]. ThunderServe further shows effective disaggregation using diverse cloud GPUs [29].

Our work explores a new point in the design space by specializing hardware for each phase in a disaggregation-based scheduling. We show that this approach substantially reduces costs while providing the same performance as existing cluster designs.

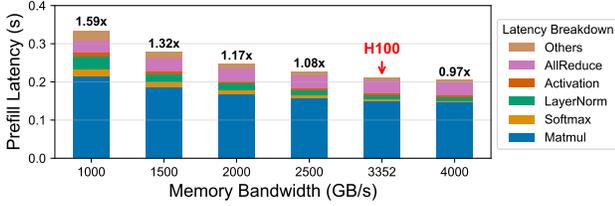

**Figure 2.** Simulated Prefill Latency Under Varying Memory Bandwidths. Hardware specifications are set according to a modeled H100 except for memory bandwidth. Simulated using LLMCompass [71] for an FP16 BLOOM-176B configuration with batch size 2, sequence length 1024, and tensor parallelism 8.

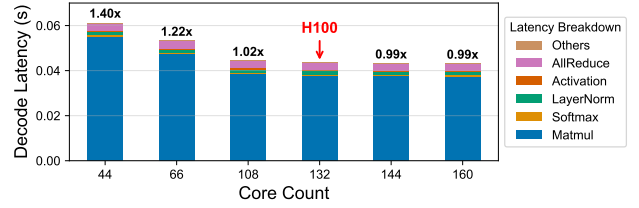

**Figure 3.** Simulated Decode Latency Under Varying Core Counts (SM count in NVIDIA GPUs). Hardware specifications are set according to a modeled H100 except the core count. Simulated by LLMCompass [71] with BLOOM-176B, batch size 64, sequence length 1024, and tensor parallelism 8.

## 3 Motivations for Phase-Specialized Hardware

Today's GPUs (also referred to as GPGPUs, General-Purpose Graphics Processing Units) are designed with a *more-is-better* philosophy to cater to the need of various workloads. This approach provides flexibility, but using the latest and greatest technologies also drives up costs. Given the dual-phase nature of LLM serving, prefill-decode disaggregation is commonly used to meet stringent latency SLOs and ensure a good user experience. In this section, we quantitatively show how GPUs are inefficient when running disaggregated prefill and decode phases. We choose BLOOM-176B [54] as a representative dense LLM because it fits on a typical 8-H100 machine, and we simulate it with FP16 precision as it can provide high production-class accuracy.

**Prefills Underutilize Memory Bandwidth.** Figure 2 shows how prefill phases underutilize the memory bandwidth on modeled H100 GPUs. We use LLMCompass [71] to simulate how the prefill phase latency changes as a function of the available memory bandwidth. Prefills are simulated with a batch size of 2 and a sequence length of 1024. We simulate an H100 GPU (3.35TB/s) as the baseline and sweep its memory bandwidth from 1TB/s to 4TB/s while keeping the rest of the hardware specifications the same.

Our results show that computation-intensive matrix multiplications dominate prefill time. Crucially, the prefill latency does not scale in proportion to the memory bandwidth. Even when the memory bandwidth is reduced to 2500 GB/s (about 0.75× that of H100), the latency only increases by 8%. This trend indicates that prefill phases do not require the large memory bandwidth provisioned on the H100 chip.

**Decodes Underutilize Compute Capacity.** Decode machines underutilize GPU compute cores when deployed using prefill-decode disaggregation due to their low arithmetic intensity. Figure 3 shows this by plotting the decode latency breakdown of BLOOM-176B while sweeping the core count (*i.e.*, the Streaming Multiprocessor count) on a simulated H100 with LLMCompass. We use FP16, a batch size of 64, a sequence length of 1024, and a tensor parallelism of 8. We vary the number of cores from 44 to 160 while setting the other hardware specifications according to a modeled H100. We find that decode performance scales sub-linearly with an increased core count. Specifically, despite using nearly 20% fewer cores (108) than an H100 (132), the decode latency only increases by about 2%. This indicates that decode phases do not require the large compute capacity provisioned on the H100 chip for efficient execution.

The prefill/decode bottleneck shifting under various conditions is further explored in Section B.1.

**Takeaway.** Underutilized hardware directly translates into increased costs for disaggregated LLM serving. These costs can be reduced by "right-sizing" regular GPU designs into separate chips to run the prefill and decode phases, respectively. The evolving models and workloads require that the hardware specialized for one phase should efficiently run the other phase to ensure flexibility [51]. In the remainder of this paper, we address these challenges and show how to tailor an existing hardware such as H100 into phase-specific designs to lower LLM serving costs at cluster scale.

## 4 SPAD: Overview

SPAD (**S**pecialized **P**refill **a**nd **D**ecode hardware) is a heterogeneous system incorporating specialized hardware for each inference phase to reduce disaggregated LLM serving costs at scale. In this section, we start by describing how SPAD clusters are organized and managed. In Section 5, we will describe the design methodology of our proposed chips.

**Cluster Organization.** Figure 4 shows an overview of a SPAD cluster. The design is similar to existing GPU-based disaggregated serving clusters but with one key difference: instead of homogeneous GPU machines, SPAD clusters consist of heterogeneous Prefill and Decode Machines optimized to run prefill and decode phases, respectively. Each proposed

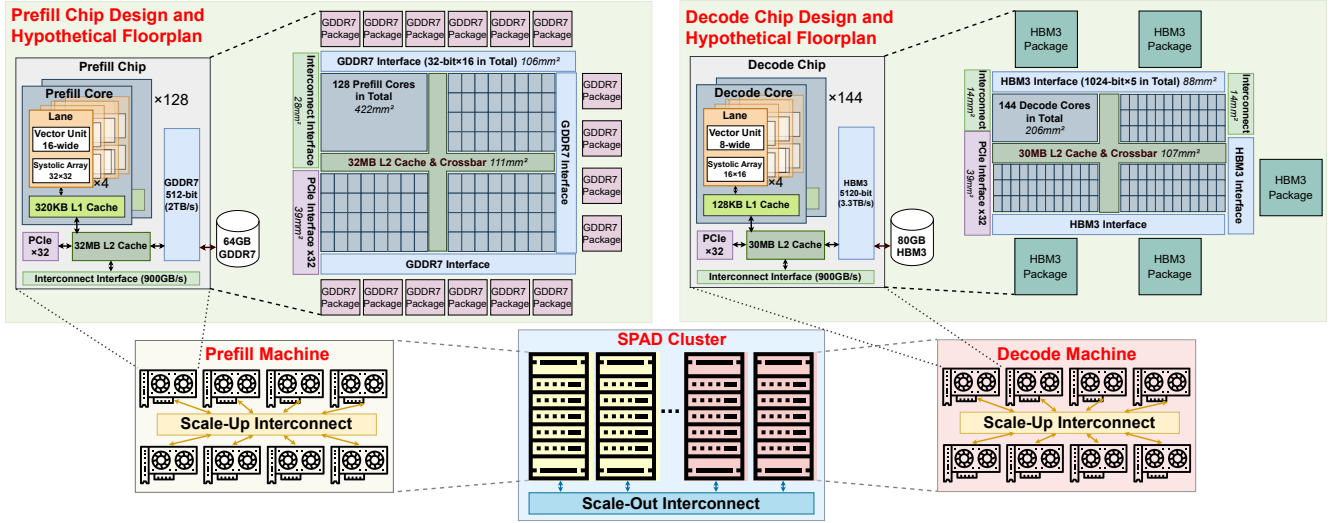

**Figure 4.** Proposed SPAD Cluster and Chips Overview. Die area is estimated and will be further explained in Section 6.1.

Prefill/Decode Machine contains 8 Prefill/Decode Chips tailored to match the computational properties of the corresponding phase. Specifically, Prefill Chips optimize compute capabilities, whereas Decode Chips optimize memory bandwidth. Chips within a machine are connected to each other with a high bandwidth scale-up interconnect (*e.g.*, NVLink). Chips across machines are connected using a lower bandwidth scale-out interconnect (*e.g.*, Infiniband). We assume the same scale-up (900 GB/s total bandwidth per chip) and scale-out (50 GB/s per chip) interconnect as NVIDIA H100s [44].

**Disaggregated Serving.** LLM replicas run separately on Prefill and Decode Machines. Incoming requests are first scheduled on Prefill Machines, which run the prefill phase and transfer the computed KV caches to Decode Machines over scale-out interconnects to finish the request [51, 74].

**Workload-Driven Provisioning.** SPAD clusters are provisioned for a target workload by deciding the number of phase-specific chips to deploy. Due to the heterogeneity of the cluster, it is necessary to carefully select the ratio of Prefill and Decode Chips to ensure optimal performance and efficiency. Given a target model and the workload distribution, cluster operators can estimate the ideal number of machines required for each phase by sweeping the cluster design space [51]. Furthermore, operators can also use existing workload estimation techniques to allocate sufficient capacity to accommodate future workload demands [11, 13].

**Adaptive Reallocation.** Models and workloads can change during the multi-year lifespan of the cluster. In such cases, the provisioned ratio of Prefill and Decode Machines may not perfectly match the new requirements, leading to suboptimal performance. Since hardware is difficult to change once deployed, SPAD clusters retain efficiency by logically reallocating Prefill and Decode Machines as needed to run either phase. This consideration is fundamentally incorporated into our hardware design methodology, which enables each chip to also run the other phase cost-effectively. Techniques to further improve adaptability are discussed in Section B.2.

## 5 SPAD: Chip Design

In this section, we describe our *less-is-more* methodology to design SPAD chips. Using the H100 GPU as a reference design, we conduct a cost-aware architectural design space exploration to tailor specialized Prefill and Decode chips for a target workload. Crucially, we design our proposed chips with the flexibility to handle either phase, enabling them to be reallocated as workload profiles evolve. Later in Section 7.2, we show the longevity of our design through adaptive reallocation.

### 5.1 *Less-is-More* Design Methodology

**Our goal is to design Prefill/Decode Chips that align with the characteristics of prefill/decode phases.** Current mainstream LLM serving hardware, such as GPUs, fails to meet this goal due to their *more-is-better* design methodology: they tend to fit as much compute capacity as is possible onto reticle-limited dies paired with high-end HBMs to provide substantial memory bandwidth and capacity. NVIDIA H100 has a die area of 814 $mm^2$ with 80 GB of HBM3 and 3.35 TB/s memory bandwidth [44]. Chiplet technology has been adopted to further increase the die area to fit more compute capacity. AMD MI300X has 8 compute chiplets with 192 GB of HBM3 and 256 MB of LLC (Last-Level Cache) [7, 8]. NVIDIA B200s are reported to have two recticle-limited dies with 186 GB of HBM3E and 8 TB/s of memory bandwidth [43]. On the other hand, Groq uses SRAM instead of HBMs, achieving a memory bandwidth of up to 80 TB/s [26].

**This *more-is-better* design methodology is not cost-effective for disaggregation-based LLM inference.** As Section 3 shows, the prefill latency of a modeled H100 only increases by 17% if we reduce the memory bandwidth by 40%, and the decode latency only increases by 22% if we cut the compute capacity by 50%. These imply that prefills do not fully utilize the memory bandwidth offered by expensive HBMs, and decodes underutilize the compute capacity of the enormous dies. Thus, we reconsider whether *more-is-better* offers a favorable trade-off between performance and cost.

**We adopt a *less-is-more* design methodology, treating cost as a first-class citizen.** Our goal is to achieve the throughput and latency requirements of LLM serving at the lowest possible cost. A large die or high-end memory can increase the manufacturing cost and TDP, and a larger TDP can lead to higher power delivery and cooling equipment costs. Thus, we assess the cost-performance trade-offs with design space explorations. If an architectural component does not significantly impact performance, we consider cutting it to save cost. However, we cannot be too aggressive with cutting components since a phase-specialized chip needs to be able to run the other phase after adaptive reallocation. By carefully choosing the memory technology and using cost-aware architectural designs, our proposed Prefill/Decode Chips could achieve similar performance with lower hardware cost and TDP than H100s since their hardware characteristics align better with the arithmetic intensity of prefill/decode phases. We call this the *less-is-more* methodology, which reduces the cost/TDP per chip while allowing more chips to be deployed in clusters under the same cost/TDP budgets to achieve higher overall performance.

### 5.2 Prefill Chip Design

**5.2.1 Memory.** Figure 2 shows that even if the memory bandwidth of a modeled H100 is reduced by 40% to 2 TB/s, its simulated prefill latency will only be 17% higher. However, further reducing the bandwidth to 1.5 TB/s increases the latency by 32%. Latency breakdowns show that the latency increase is mainly caused by memory-bound non-tensor operations, such as Layer Normalization or Softmax, whose performance scales almost linearly with memory bandwidth. On the other hand, Matmul latency only increases by 16%, even when decreasing memory bandwidth from 4 TB/s to 2 TB/s. Therefore, we conclude that reduced memory bandwidth can be manageable as long as it is not decreased too much, and the increase in non-tensor operation latency can be compensated for by increasing Matmul performance. Also, since Prefill Machines only temporarily store the KV cache before transferring it to the Decode Machines, their memory capacity requirement is lower than that of the decode phase.

**Following our *less-is-more* design philosophy, we propose to replace HBMs with GDDR memory as a cheaper alternative for prefill**, which is commonly used in gaming GPUs [46] and desktop workstation GPUs [41].

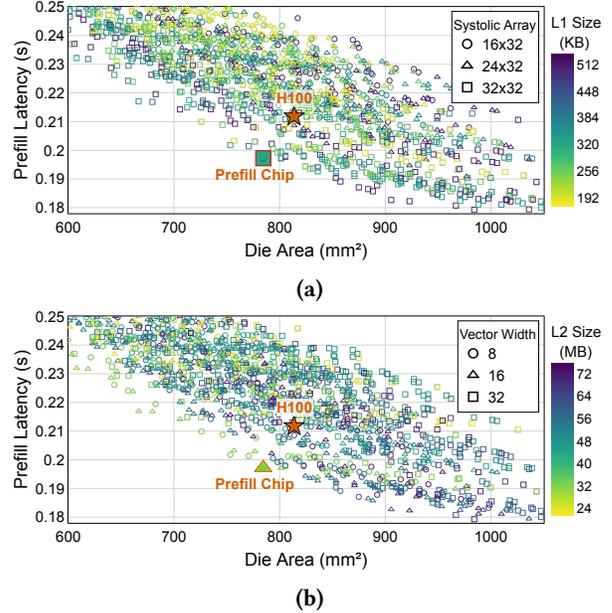

**Figure 5.** Prefill Chip Design Space Exploration. Latencies of our chips and H100 are all simulated with LLMCompass [71]. Die areas of our chips are estimated and will be further explained in Section 6.1. H100 die area is reported by NVIDIA [44]. We use FP16 BLOOM-176B with tensor parallelism 8, sequence length 1024, and batch size 2. Larger systolic arrays significantly boost prefill performance. Smaller vector units have minimal performance impact.

**Table 2.** Comparison of Memory Technologies

|  | Source Processor | Processor Bandwidth | Bandwidth/ Beachfront(PHY)♦ | Estimated Cost♠ |
|---|---|---|---|---|
| LPDDR5X | Apple M4 (3 nm) | 120 GB/s | 8 GB/s/mm | ? |
| GDDR7 | RTX 5090 (4 nm) | 1792 GB/s | 22 GB/s/mm | $3/GB |
| HBM3 | H100 (4 nm) | 3352 GB/s | 68 GB/s/mm | $9/GB |

♦ Bandwidth per beachfront of the processor PHYs, estimated based on specifications and annotated die photos [36, 44, 46, 56, 63].
♠ Cost modeling is explained in Section 6.1.

We do not choose other memory technologies like LPDDR since they have a lower bandwidth under the chip beachfront limits and do not meet the requirements of the prefill phase, as shown in Table 2. In contrast, as Table 3 shows, GDDR7 can provide 2 TB/s of bandwidth and 64 GB of capacity with a 512-bit bus and 16 packages, which meets prefill requirements. We estimate substituting HBM with GDDR could reduce memory cost by 3×, which we further explain in Section 6.1. Note that our proposed Prefill Chip has a smaller memory capacity (64 GB) compared to H100s (80 GB), and later in Section 7 our end-to-end simulations show that it is not a bottleneck for prefill phases since the KV cache is only temporarily stored.

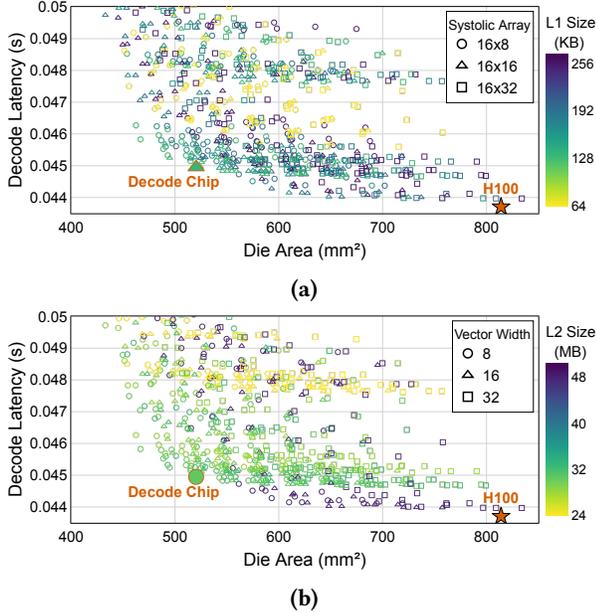

**Figure 6.** Decode Chip Design Space Exploration. Latencies of our chips and H100 are all simulated with LLMCompass [71]. Die areas are estimated and will be explained in Section 6.1. H100 die area is reported by NVIDIA [44]. We use FP16 BLOOM-176B with tensor parallelism 8, sequence length 1024, and batch size 64. Our design strikes a desirable balance between performance and die area.

**Table 3.** SPAD Chips Compared with NVIDIA H100

| Specifications | Prefill Chip | Decode Chip | H100 [44] |
|---|---|---|---|
| Core Count | 128 | 144 | 132 |
| Lane per Core | 4 | 4 | 4 |
| Vector Width | 16 | 8 | Eq. to 32 |
| Systolic Array | 32 × 32 | 16 × 16 | Eq. to 16 × 32 |
| L1 Cache per Core | 320 KB | 128 KB | 256 KB |
| L2 Cache | 32 MB | 30 MB | 50 MB |
| Memory Protocol | GDDR7 | HBM3 | HBM3 |
| Memory Bus Width | 512-bit | 5120-bit | 5120-bit |
| Pin Speed | 32 Gb/s | 5.2 Gb/s | 5.2 Gb/s |
| Memory Package Count | 16 | 5 | 5 |
| Capacity per Package | 4 GB | 16 GB | 16 GB |
| Clock (Tensor) | 1.83 GHz | 1.83 GHz | 1.83 GHz |
| Clock (Non-Tensor) | 1.98 GHz | 1.98 GHz | 1.98 GHz |
| FP16/BF16 Tensor PFLOPs | 1.92 | 0.54 | 0.99 |
| FP32 Non-Tensor TFLOPs | 32.4 | 18.2 | 66.9 |
| Total L1 & L2 Cache Size | 73 MB | 48 MB | 84 MB |
| Memory Configuration | 64 GB GDDR7 | 80 GB HBM3 | 80 GB HBM3 |
| Memory Bandwidth | 2048 GB/s | 3352 GB/s | 3352 GB/s |
| Est. Die Area (@4nm)♦ | 784 $mm^2$ | 520 $mm^2$ | 814 $mm^2$ |
| Est. Die Cost♦ | $301 | $187 | $315 |
| Est. Memory Cost♦ | $192 | $720 | $720 |
| Est. Norm. Total HW Cost♦ | 0.48 | 0.88 | 1 |
| Est. TDP♦ | 596 W | 507 W | 700 W |
| Norm. Prefill Perf.✿ | 1.08 | 0.69 | 1 |
| Norm. Decode Perf.✿ | 0.80 | 0.97 | 1 |

♦ The die area for our Prefill/Decode Chip is estimated. H100 die area is reported by NVIDIA [44]. Cost and TDP modeling is explained in Section 6.1.
✿ Performance numbers are from Fig. 7, simulated with LLMCompass [71].

**5.2.2 Compute.** Figure 1 shows the compute-intensive nature of the prefill phase. We observe that tensor operations, such as Matmuls, largely contribute to the compute intensity of prefill phases, and are commonly mapped to systolic arrays (or Tensor Cores in NVIDIA GPUs). On the other hand, memory-bound non-tensor operations, such as Layer Normalization, are mapped to general-purpose vector units (or CUDA Cores in NVIDIA GPUs).

**Consequently, we propose to increase the tensor compute capacity to accelerate compute-bound tensor operations and reduce the non-tensor compute capacity since those non-tensor operations are memory-bound anyway.** As shown in Figure 5, we use LLMCompass [71] to conduct a design space exploration for different combinations of core counts, vector widths, systolic array sizes, and cache sizes. We find that increasing the size of the systolic arrays significantly increases prefill performance while decreasing the size of the vector units has minimal performance impact. The L1 cache size is increased to accommodate the larger systolic arrays. We also decrease the L2 cache size as we find that approximately 30 MB of L2 is enough for LLM inference, and there is a diminishing return in further increasing L2 sizes.

### 5.3 Decode Chip Design

**5.3.1 Memory.** According to Figure 1, the decode phase is heavily memory bandwidth bound. While Prefill Machines only temporarily retain KV caches before transferring them, Decode Machines retain and use the KV cache for the rest of the request processing. Specifically, multiple requests are often batched together in decode phases to improve model weight reuse, and we need to store the KV cache for all of them. Also, since every newly generated token contributes to the KV cache, the KV cache size grows continuously until the request is complete. Therefore, the decode phase has higher memory capacity requirements to store these KV caches. Based on these observations, we choose to use HBM3 due to its high bandwidth and large capacity. Unlike Groq [26], we do not consider on-chip SRAM due to the high cost it would require to meet the memory capacity requirement.

**5.3.2 Compute.** Figure 3 shows that the simulated decode latency only increases by 22% even if we reduce the core count of a modeled H100 by half. Due to the *more-is-better* design methodology of H100s, the compute capacity remains underutilized for decodes. To eliminate this inefficiency, we follow our *less-is-more* philosophy and conduct a design space exploration to sweep different architectural configurations, as shown in Figure 6, identifying which hardware

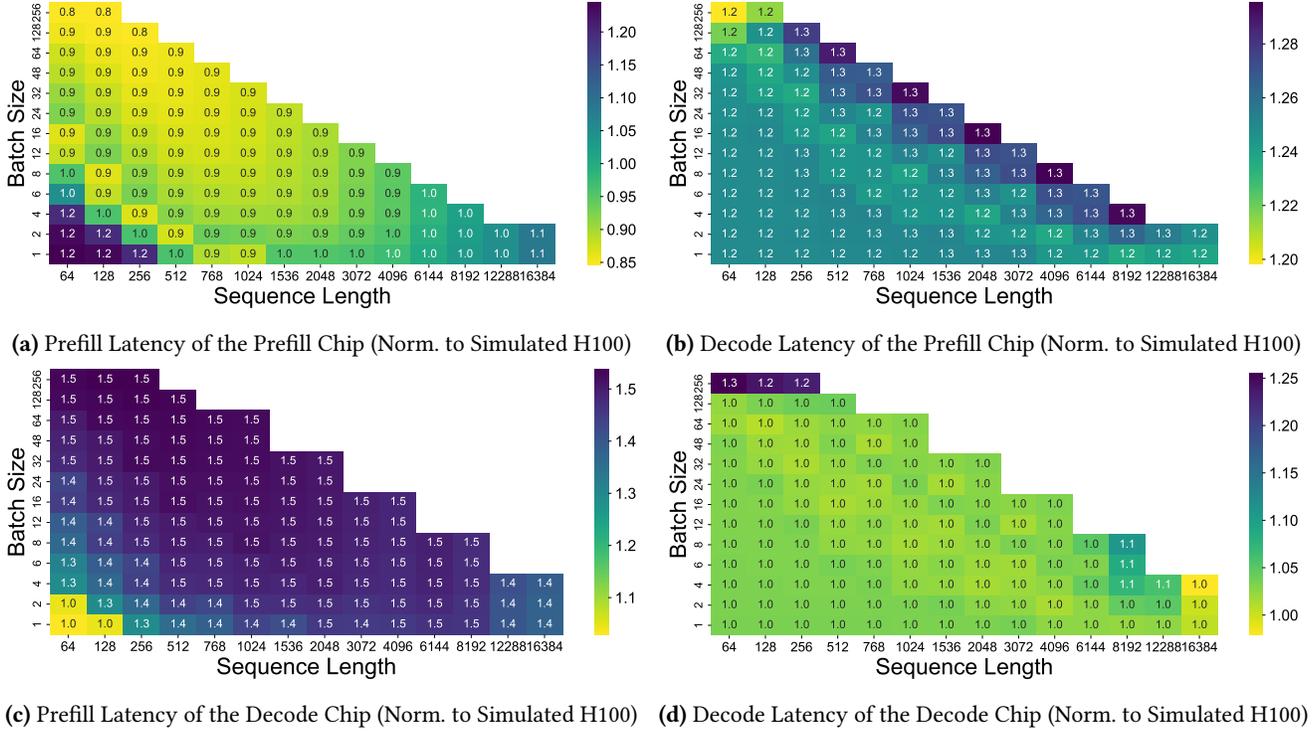

**Figure 7.** Chip Performance Under Various Batch Sizes and Sequence Lengths. Lower is better. Latencies of our proposed chips and H100 are all simulated with LLMCompass [71] modeling FP16 BLOOM-176B with tensor parallelism 8. Only the combinations that fit the memory capacity are shown. A sensitivity study on parallelism strategies is further shown in Fig. 11 in Section A.

resources and to what extent can be cut without impacting performance.

**We find that smaller systolic arrays and vector units are more efficient than larger ones for decode.** Due to the low arithmetic intensity and memory-bound nature, large systolic arrays and vector units bring very marginal performance gains since decode phases cannot fully utilize them. Therefore, our proposed Decode Chip adopts a systolic array size of 16×16 and a vector width of 8. We did not further reduce tensor performance because the area savings were outweighed by the significant slowdown of running prefill, which can affect flexibility after reallocation.

**We find that smaller caches are sufficient for decode.** Large caches improve performance through better memory reuse. However, since decode phases are memory-bound in reading model weights and KV caches, larger caches do not help much for these streaming memory accesses. Compared to a modeled H100, we cut the L1 size by 50% and the L2 size by 40%.

### 5.4 Summary

A comparison of our proposed Prefill/Decode Chips with H100s is summarized in Table 3. The extra complexity arising from heterogeneous chips is discussed in Section B.3.

**Prefill Chip.** Compared to a modeled H100, our proposed Prefill Chip roughly doubles the tensor performance while maintaining a similar die area by reducing the unessential non-tensor performance by half and cutting down L2 cache size. A hypothetical floorplan is shown in Figure 4.

Figures 7a and 7b show the performance of the proposed Prefill Chip simulated with LLMCompass [71]. Compared to a modeled H100, it is 8% faster for prefills on average: the tensor operations are faster due to larger systolic arrays, but the non-tensor operations are slower due to reduced memory bandwidth. The hardware cost is reduced by 52%, mainly from substituting HBMs with cheaper GDDR7 memory.

Our proposed Prefill Chip can be slightly slower than modeled H100s on very few total batched tokens (≤ 256 tokens) or very long input prompts (≥ 12288 tokens). Very short input sequences have little weights reuse and low arithmetic intensity. For very long input sequences, the Softmax operation inside the attention mechanism becomes more dominant due to its quadratic complexity with respect to sequence length. Since our Prefill Chip has less memory bandwidth and non-tensor compute capability, Softmax becomes the new bottleneck. This bottleneck could be alleviated by chunking long prefills [4] or using sequence parallelism [64].

**Decode Chip.** The key specifications of our proposed Decode Chip are summarized in Table 3, and a hypothetical floorplan is shown in Figure 4. Compared to a modeled H100, our proposed Decode Chip reduces the die area by 36% and lower the TDP by 28%, primarily due to its lower compute capacity and smaller caches. Figure 7d shows that it still achieves 97% of the decode performance of a modeled H100 on average. Our Decode Chip can be slower for very large batch sizes (≥ 256) due to increased arithmetic intensities. However, such large batch sizes can be rare in production, especially for latency-sensitive workloads, due to the HBM capacity limit and the diminishing return of batching.

## 6 Evaluation Methodology

### 6.1 Cost and TDP Modeling

**Total Hardware Cost.** We account for the combined manufacturing costs of the die and memory. We exclude other costs such as masking, packaging, and design, since they are not commonly disclosed and estimates vary widely. Additionally, when manufacturing at scale, one-time mask and design costs can be amortized across all dies.

We modify LLMCompass' area model to model the die areas of our proposed Prefill Chip and Decode Chip guided by annotated H100 die photos [36]. For our proposed designs, we assume a 10% area overhead to account for white space and disabled defective components and a TSMC 4nm process node. We assume 4nm wafer costs of $20,000 per 300mm wafer, which aligns with estimates for modern process nodes [39, 61, 67]. To find die costs, we calculate the number of dies that can fit a single wafer.

For device memory costs, we estimate $3 per GB for GDDR7 based on current GDDR6 spot prices [60]. HBM pricing is less transparent, and estimates vary between $10 to $35 per GB [10, 21]. For our cost model, we assume that HBM costs are between 2×-4× the cost of GDDR based on publicly disclosed industry estimates [23]. In Section A Table 9, we further explore different HBM3 cost assumptions. Note that even the highest 1:4 ratio of $12 per GB is on the lower end of cost estimates for HBM3.

Table 3 details die area estimates for the proposed Prefill/Decode Chips. Based on these, we calculate die costs, memory costs (assuming a 1:3 GDDR7:HBM3 cost ratio), and total hardware costs for the three devices.

**TDP Modeling for our Prefill/Decode Chip.** The H100 has a TDP of 700 W [44], and we assume a 10% TDP overhead to account for VRM conversion loss and other peripherals [5, 35]. We assume each HBM package has a power consumption of 30 W [57]. Based on these, we assume the H100 die itself excluding HBMs has a TDP of $700 \times 90\% - 30 \times 5 = 480W$, and we assume our Prefill/Decode Chip has the same power density as an H100 die. GDDR7 power consumption is estimated by the reported 4.5 pJ/bit from Micron [37].

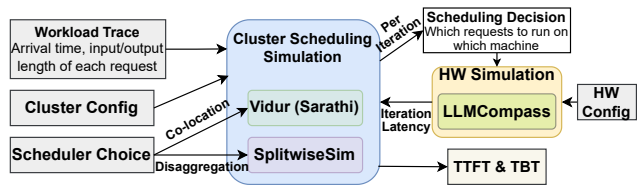

**Figure 8.** End-to-End Simulation Setup

### 6.2 End-to-end Simulations

We perform end-to-end evaluations that include both hardware architectural simulation and cluster-level scheduling simulation with workload traces to estimate how our designs translate into performance and cost improvements at scale. An overview of our simulation setup is shown in Figure 8.

Given a cluster configuration and a workload trace, the scheduler will dispatch each request to a machine within the cluster. We explore two such scheduling approaches: SplitwiseSim [51] as an implementation for disaggregated scheduling (Splitwise), and Vidur [3] as an implementation for co-location-based scheduling (Sarathi [4]). The fidelity of these scheduler implementations has been explored in their corresponding publications [3, 51].

At each iteration, cluster simulators make scheduling decisions to assign requests to machines, and these iteration-level request batches are fed into LLMCompass [71] to estimate how long it will take each machine to finish one iteration for the request batch scheduled upon it.

We extended LLMCompass to support H100 modeling and new models with a similar error rate as in the original paper [71]. We also extended SpitwiseSim and Vidur to use LLMCompass as their performance model. With LLMCompass serving as the unified architectural performance model across different schedulers and hardware, fair comparisons are achieved. **All results in this paper are simulated, not measured on real hardware.** In other words, we model the execution rather than performing the computation with actual parameter values.

### 6.3 Experimental Setup

**Models.** We evaluated three open-source models with different sizes, model architectures, and deployment strategies: ① BLOOM-176B [54] uses Multi-Head Attention [62] and we deployed it with FP16 and a tensor parallelism of 8 (TP=8). ② Llama3-70B [25] uses Grouped-Query Attention [6] with smaller KV cache footprints, and we deployed it with FP16 and TP=4. ③ DeepSeek-V2-236B [19] uses DeepSeekMoE [17] with Multi-head Latent Attention to compress the KV cache, and we deployed it with FP8 and expert parallelism 8 (EP=8).

**Workloads.** We use open-source request traces from Microsoft [12], representing two common LLM applications: *coding* (code completion) and *conversation* (chatbot). The

**Table 4.** Provisioning Results Summary

|  | Coding (70 req/s) | | | Conversation (70 req/s) | | |
| --- | --- | --- | --- | --- | --- | --- |
|  | HW Requirement✻ | Norm. HW Cost◆ | Norm. TDP◆ | HW Requirement✻ | Norm. HW Cost◆ | Norm. TDP◆ |
| Sarathi | 36 H100 | 36 | 36 | 34 H100 | 34 | 34 |
| Splitwise-homo | 25 H100 | 25 | 25 | 23 H100 | 23 | 23 |
| Splitwise-hetero♠ | 21 H100 + 9 A100 | 25.5 | 25.5 | 13 H100 + 32 A100 | 29 | 29 |
| Splitwise-pcap | 21 H100 + 4 450W H100 | 25 | 23.6 | 6 H100 + 21 450W H100 | 27 | 19.5 |
| **SPAD** (P+D) | 18 Prefill + 7 Decode | **14.7** | **20.4** | 8 Prefill + 17 Decode | **18.7** | **19.1** |

✻ Minimum number of modeled 8-chip machines to meet the SLOs with BLOOM-176B. The unit is 8-chip machines, e.g., 36 H100 refers to 36 modeled 8-H100 machines and 18 Prefill refers to 18 8-Prefill-Chip machines. ◆ Normalized to the HW cost/TDP of a modeled 8-H100 machine. ♠ Assumes that A100s have half the hardware cost and TDP of H100s.

**Table 5.** Latency SLOs. Defined as the slowdown relative to running the request on modeled H100s without batching.

| SLOs✻ | P90 TBT | P90 TTFT | P99 TBT | P99 TTFT |
| --- | --- | --- | --- | --- |
| Loose/**Normal**/Tight | 2.5×/**2**×/1.5× | 4×/**3**×/2× | 6×/**5**×/3× | 8×/**6**×/4× |

✻ Normal SLOs are used unless otherwise specified.

coding workload has long input prompts (median: 1500 tokens) and short output sequences (median: 13 tokens), while the conversation workload has shorter input prompts (median: 1020 tokens) and longer output sequences (median: 129 tokens).

**SLOs.** We evaluate the maximum throughput that can be supported under normalized P90 and P99 TTFT and TBT SLOs, shown in Table 5. Similar to prior work [51], SLOs are defined relative to the execution latencies of the same request without any batching or contention on modeled H100s.

**Baselines.** We use Splitwise [51] and Sarathi [4] as baseline GPU-driven LLM serving cluster systems. Splitwise is a disaggregation-based system, and we compare SPAD with three of its variants: Splitwise-homo with H100s, Splitwise-pcap that uses H100s for prefill and hypothetical power-capped H100s (450W TDP)[1] for decode, and Splitwise-hetero with H100s for prefill and A100s for decode. Sarathi is a colocation-based system, and we configure it with modeled H100s. All baselines are evaluated as described in Section 6.2.

## 7 Results

### 7.1 Cluster Provisioning

We start by evaluating the efficacy of SPAD clusters when provisioned for a specific workload. Table 4 summarizes the provisioning results for BLOOM-176B using the coding and conversation workloads with a target request rate of 70 req/s.

[1]In order to have a baseline that optimizes for TDP, we assume a hypothetical power-capped 450 W TDP H100 for decode with 76% of the peak FP16 tensor TFLOPs while retaining the same memory and interconnect specifications as the original 700 W H100. Since we do not have access to H100 power configurations and its dynamic voltage and frequency scaling (DVFS) implementation, our LLMCompass simulation is based on the hardware specification of the 350 W NVIDIA H100 PCIe [44]. We substitute its low-end memory and NVLink interconnect with those of the original 700W H100, which we assume adds 100 W TDP.

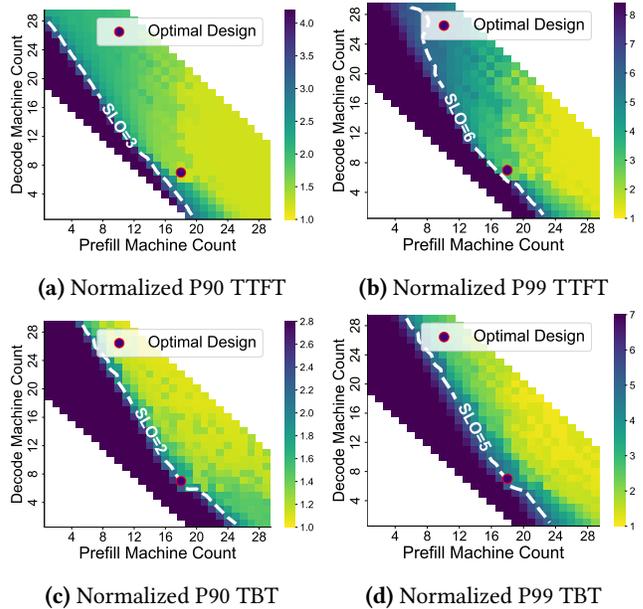

(a) Normalized P90 TTFT  (b) Normalized P99 TTFT

(c) Normalized P90 TBT   (d) Normalized P99 TBT

**Figure 9.** Provisioning Results with Coding Trace for SPAD. The optimal design has 18 prefill and 7 decode machines.

**Coding.** Compared to the best baseline, SPAD saves the hardware cost by **41%** and TDP by **13%**. Sarathi needs at least 36 modeled 8-H100 machines to meet all SLOs, while Splitwise needs at least 25 modeled 8-H100 machines. The minimal hardware requirement is derived by sweeping machine count as shown in Figures 12 and 13a in Section A. Due to prefill-decode interference, Sarathi can be unsuitable for low-latency workloads compared to disaggregated serving. Splitwise-hetero needs at least 21 modeled 8-H100 and 9 modeled 8-A100 machines, which does not improve cost-effectiveness because although the TFLOPS-to-memory-bandwidth ratio of A100 is closer to the theoretical arithmetic intensity of decode, the absolute bandwidth and TFLOPS are significantly lower, making it harder to meet strict latency SLOs. Figure 9 shows that SPAD needs the same amount of

**Table 6.** Provisioning Results under Various SLOs.

| Workloads | Coding (70 req/s) | | | Conversation (70 req/s) | | |
|---|---|---|---|---|---|---|
| SLOs | Loose | Normal | Tight | Loose | Normal | Tight |
| Sarathi (H100)♦ | 33 | 36 | 45 | 31 | 34 | 40 |
| Splitwise (H100)♦ | 24 | 25 | 27 | <u>22</u> | <u>23</u> | <u>27</u> |
| Splitwise (H100+A100)♦ | 20+9 | 21+9 | 27+0 | 13+20 | 13+32 | 27+0 |
| Splitwise (H100+pcap)♦ | <u>19+5</u> | <u>21+4</u> | <u>23+4</u> | 3+23 | 6+21 | 11+23 |
| SPAD (P+D)♦ | 18+6 | 18+7 | 21+7 | 8+17 | 8+17 | 13+14 |
| Hardware Saving✤ | 42% | 41% | 40% | 15% \| 28% | 19% \| 31% | 32% \| 46% |
| TDP Saving✤ | 11% | 13% | 10% | 13% \| -8% | 17% \| 2% | 21% \| 18% |

♦ The unit is modeled 8-chip machines.
✤ Hardware/TDP saving of SPAD compared to pareto-optimal baselines (underlined). There can be more than one Pareto-optimal baseline: Splitwise (H100) has lower hardware cost but higher TDP than Splitwise (H100+pcap).

**Table 7.** SPAD Reallocation After Changing Workload (Model Remains Unchanged: BLOOM-176B)

| Provisioned Cluster (P+D)✤ | Reallocated Workload | Reallocated Throughput | Min. HW✤♦ for Splitwise | (HW, TDP) Saving |
|---|---|---|---|---|
| 18P+7D♠ | Conversation | 55 req/s | 19 H100 | (23%, -7%) |
| 8P+17D♣ | Coding | 60 req/s | 21 H100 | (11%, 9%) |

✤ The unit here is modeled 8-chip machines.
♦ The minimum hardware required to achieve the reallocated throughput. Sarathi performs consistently worse than Splitwise and is not shown.
♠ Initially provisioned for Coding (70 req/s).
♣ Initially provisioned for Conversation (70 req/s).

machines as Splitwise-homo, demonstrating the effectiveness of our *less-is-more* design methodology.

**Conversation.** Compared to the two Pareto-optimal baselines, SPAD saves hardware cost and TDP by **(19%, 17%)** relative to Splitwise-homo, and by **(31%, 2%)** relative to Splitwise-pcap. Splitwise-pcap saves TDP, but not hardware cost. Figure 14 in Section A shows the detailed provisioning results for SPAD. Compared to the coding workload, the conversation workload requires more Decode Machines due to its longer output sequences.

**Changing SLOs.** Table 6 shows the minimum hardware requirements to sustain 70 req/s under three sets of SLOs from loose to tight (as defined in Table 5), demonstrating SPAD's consistent performance under various SLOs. The Normal SLOs are the ones used in previous experiments.

### 7.2 Cluster Reallocation

Next, we evaluate how well an already provisioned SPAD cluster performs after reallocation when the workloads and models change. In this section, we compare with the modeled 700W TDP H100 as the baseline hardware due to its balanced performance across various workloads and SLO settings in the provisioning experiments.

**Changing Workloads.** Table 7 and Fig. 10a show that the cluster that was initially provisioned for the coding workload at 70 req/s can be repurposed for the conversation workload

**Table 8.** SPAD Reallocation After Changing the Model (Workload Remains Unchanged)

| Provisioned Cluster (P+D)✤ | Reallocated Model | Reallocated Throughput | Min. HW✤♦ for Splitwise | (HW, TDP) Saving |
|---|---|---|---|---|
| 18P+7D♠ | Llama3-70B | 188 req/s | 26 H100 | (43%, 22%) |
| 8P+17D♣ | Llama3-70B | 171 req/s | 27 H100 | (31%, 29%) |
| 18P+7D♠ | DeepSeek-V2 | 103 req/s | 23 H100 | (36%, 11%) |
| 8P+17D♣ | DeepSeek-V2 | 183 req/s | 24 H100 | (22%, 20%) |

✤ The unit here is modeled 8-chip machines.
♦ The minimum hardware required to achieve the reallocated throughput.
♠ Initially provisioned for Coding (70 req/s) with BLOOM-176B.
♣ Initially provisioned for Conversation (70 req/s) with BLOOM-176B.

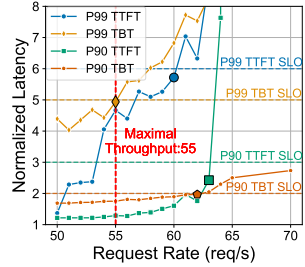
(a) Coding-Opt Cluster Running Conversation (BLOOM-176B)

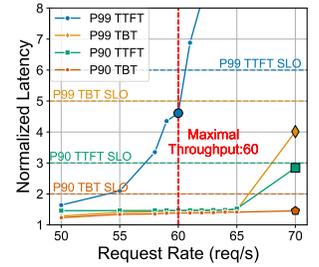
(b) Conversation-Opt. Cluster Running Coding (BLOOM-176B)

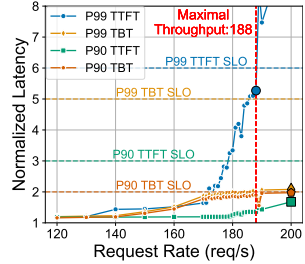
(c) BLOOM-Opt. Cluster Running Llama3-70B (Coding)

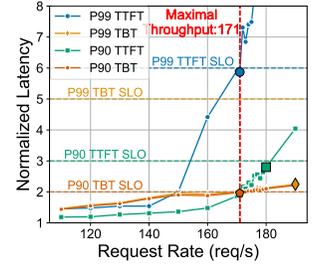
(d) BLOOM-Opt. Cluster Running Llama3-70B (Conversation)

**Figure 10.** SPAD Clusters After Reallocation. Markers indicate the highest feasible request rate under each SLO; their minimum is the maximum supported cluster throughput.

at 55 req/s after reallocation, where 8 Prefill Machines are reallocated to run decode. The baseline needs at least 19 modeled 8-H100 machines to achieve the same throughput, so SPAD can still save the hardware cost by **23%** at the cost of **7%** larger TDP. Although the Prefill Chip runs decode at a reduced hardware efficiency, its hardware cost saving by using GDDR instead of HBM is still significant.

Table 7 and Fig. 10b show that the cluster initially provisioned for the conservation workload at 70 req/s can be reallocated to support 60 req/s for the coding workload, where 14 Decode Machines are reallocated for prefill. The baseline needs at least 21 modeled 8-H100 machines to achieve the same throughput, so SPAD still reduces the hardware cost by **11%** and TDP by **9%**. We attribute this benefit to the fact that our Decode Chip is designed to run prefill phases reasonably well, so it does not sacrifice the performance much.

**Changing Models.** Table 8 shows that when the model evolves from Multi-Head Attention (MHA) to Grouped-Query Attention (GQA) and Multi-head Latent Attention (MLA) and MoE (DeepSeek-V2), the cluster initially provisioned for BLOOM-176B can also serve Llama3-70B and DeepSeek-V2 efficiently, achieving **22%-43%** hardware cost saving and **11%-29%** TDP saving compared to the modeled H100 baseline.

For Llama3-70B, the cost savings tend to be greater than for running BLOOM-176B, mainly because GQA enables Key/Value sharing inside each group, which increases the arithmetic intensity and favors our proposed Prefill Chips. For DeepSeek-V2, the cost savings are smaller mainly because it is a sparse MoE model and has smaller arithmetic intensity. In DeepSeek-V2, tokens are dispatched to 160 different routed experts. Since each token only activates a subset of total weights, the per-expert weight reuse across different tokens is smaller compared to dense models.

## 8 Conclusion

This work introduces SPAD, a heterogeneous system to accelerate disaggregation-based LLM serving. Leveraging the dual-phase nature of LLM inference, we adopt a *less-is-more* philosophy to design cost-effective Prefill and Decode Chips tailored to their distinct computational characteristics. Compared to modeled H100s, our proposed Prefill Chips deliver 8% higher prefill performance on average at 52% lower hardware cost, while our proposed Decode Chips achieve 97% of the decode performance with 28% lower TDP. End-to-end simulations show that SPAD reduces hardware costs by 19%-41% and TDP by 2%-17% compared to modeled baseline clusters while maintaining the same performance. As models and workloads change, SPAD can perform an adaptive chip reallocation and still achieve 11%-43% lower hardware costs, demonstrating the longevity of our design.

## Acknowledgments

This work was supported in part by ACE, one of the seven centers in JUMP 2.0, a Semiconductor Research Corporation (SRC) program sponsored by DARPA. This material is based upon work supported by a Princeton Andlinger Center Innovation Award, a Princeton SEAS Innovation Award, and the National Science Foundation Graduate Research Fellowship Program under Grant No. DGE-2039656. Any opinions, findings, and conclusions or recommendations expressed in this material are those of the author(s) and do not necessarily reflect the views of the National Science Foundation. This work was also supported by the Princeton Yan Huo *94 Graduate Fellowship.

## A  Supplementary Results

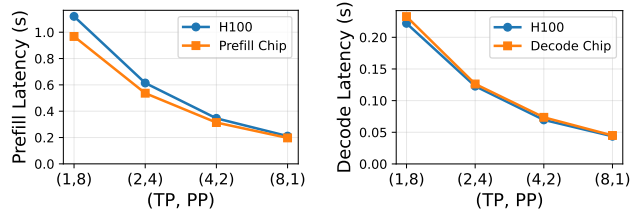

**Figure 11.** Chip Performance Under Various Tensor (TP) and Pipeline Parallelism (PP). Latencies of our chips and H100 are all simulated with LLMCompass [71] and FP16 BLOOM-176B with sequence length 1024 and batch size 2 and 64 for prefill and decode respectively. Our proposed chips perform consistently under various model parallelisms.

**Table 9.** Chip Cost under Various HBM Cost Assumptions

| HBM Cost Assumptions | $6/GB | $9/GB✤ | $12/GB |
|---|---|---|---|
| Estimated HBM Cost | $480 | $720 | $960 |
| Estimated Decode Chip Cost | $667 | $907 | $1147 |
| Estimated H100 Cost | $795 | $1035 | $1275 |

✤ We use $9/GB for HBM cost in the the paper.

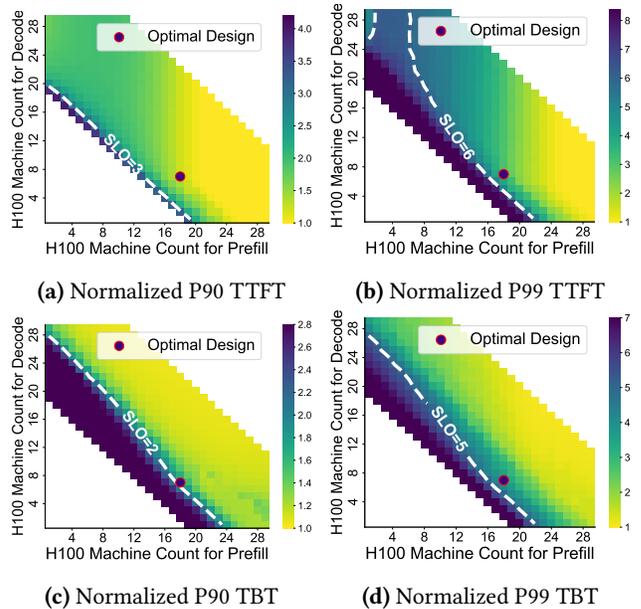

(a) Normalized P90 TTFT    (b) Normalized P99 TTFT

(c) Normalized P90 TBT    (d) Normalized P99 TBT

**Figure 12.** Provisioning Results with Coding Trace (70 req/s) and BLOOM-176B for Splitwise-homo. At least 25 modeled 8-H100 machines are required to meet all the SLOs. Markers indicate one of the Pareto-optimal designs with 18 modeled 8-H100 machines for prefill and 7 for decode. All results here are simulated as explained in Section 6.2.

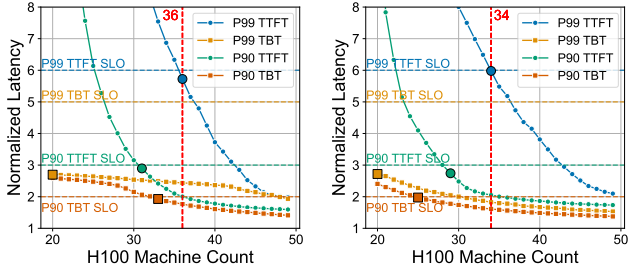

(a) Coding (70 req/s): 36 H100 machines to meet all SLOs
(b) Conversation (70 req/s): 34 H100 machines to meet all SLOs

**Figure 13.** Provisioning Results with Sarathi (BLOOM-176B). Markers indicate the minimum modeled 8-H100 machine count to meet each SLO. All results here are simulated as explained in Section 6.2.

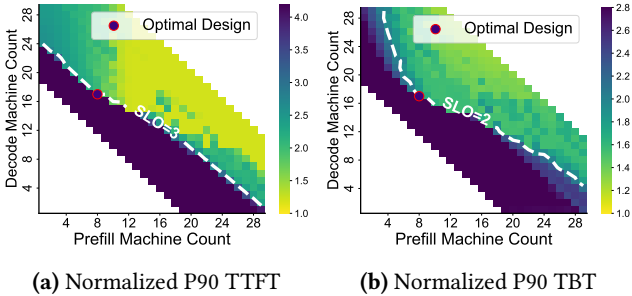

(a) Normalized P90 TTFT
(b) Normalized P90 TBT

**Figure 14.** Provisioning Results with Conversation Trace for SPAD. The optimal design has 8 prefill and 17 decode machines. P99 TTFT/TBT figures are similar and not shown.

## B Extended Discussion

### B.1 Prefill/Decode Bottleneck Shifting

In Section 3, we show that prefill is compute-bound and decode is memory-bandwidth-bound under a common batch size and sequence length setting. However, these bottlenecks can dynamically shift under various conditions:

**Prefill with very short sequences can shift towards memory-bandwidth-bound due to limited data reuse.** Fig. 15a shows that the prefill latency is more sensitive to memory bandwidth when the sequence length is very small (e.g., 64). As a result, on the bottom left corner in Fig. 7a, our proposed Prefill Chip can be slower than the modeled H100s when the batched token size is small.

**Prefill with long sequences can shift towards memory-bound.** The quadratic complexity of the attention puts more pressure on memory. Fig. 15a shows that the prefill latency is more sensitive to memory bandwidth with long sequences. On the bottom right in Fig. 7a, the performance improvement of our proposed Prefill Chip diminishes and eventually reverses. Memory capacity becomes another bottleneck with

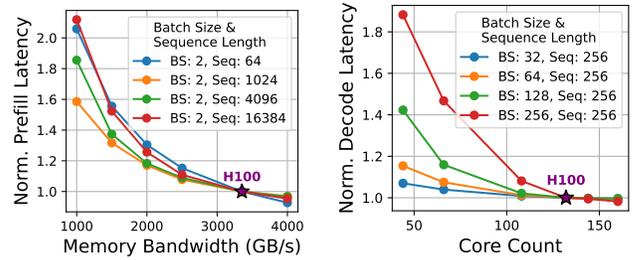

(a) Normalized Prefill Latency. Hardware specifications are set according to a modeled H100 except for memory bandwidth.
(b) Normalized Decode Latency. Hardware specifications are set according to a modeled H100 except the core count.

**Figure 15.** Prefill/Decode Latency Under Various Settings. Simulated using LLMCompass [71] for an FP16 BLOOM-176B with tensor parallelism 8. All results are normalized to simulated H100s. (a) Prefill shifts towards memory-bandwidth-bound under very long or short sequences. (b) Decode shifts towards compute-bound under large batch sizes.

long sequences due to increasing KV cache size: For FP16 BLOOM-176B, assuming 90% of the memory capacity reserved for model weights and KV cache, 8 of our proposed Prefill Chips (64GB each) can store roughly 35K tokens, while 8 modeled H100s (80GB each) can store around 66K tokens.

**Decode with large batch sizes can shift towards compute-bound.** Fig. 15b shows that decode can be more sensitive to compute capacity with large batch sizes due to increased arithmetic intensity. On the top left corner of Fig. 7d, our proposed Decode Chips are slower than modeled H100s under batch size 256. However, this condition can be rare due to KV cache sizes and latency constraints.

### B.2 Adaptability to Highly Variable Workloads

As shown in Section 7.2 and Tables 7 and 8, we rely on cluster reallocation to repurpose our proposed chips when the Prefill-to-Decode ratio of the workload changes dramatically. To further improve adaptability and robustness, we provide two recommendations:

**Buffer pool.** SPAD can be combined with a buffer pool composed of existing balanced hardware such as NVIDIA H100s. When the prefill and decode demands change, different portions of this pool can be allocated to prefill and decode according to the changing demands dynamically. We envision most of the workload still served by our proposed chips for the hardware cost and TDP benefits, with the buffer pool mainly to account for future workload variability.

**Load predictor.** At the orchestrator level, SPAD can be further combined with a runtime load predictor such as ARIMA (Autoregressive Integrated Moving Average) or Meta

Prophet [59], which has been incorporated in industry frameworks such as NVIDIA Dynamo Planner [1, 45]. At each time interval, the load predictor estimates the prefill loads and decode loads, which can be used to guide SPAD cluster reallocation under highly variable workloads.

### B.3 Extra Complexity of Heterogeneous Chips

The *less-is-more* design methodology illustrates how to take an existing LLM serving hardware and tailor it into two specialized chips for different phases. In this design process, the baseline design and the derived prefill/decode chips share architectural similarity, minimizing the compatibility issue with existing software stacks and alleviating the extra NRE and software implementation costs. For example, the different systolic array sizes between the prefill and decode chips may require tuning tiling parameters, rather than developing two entirely different software implementations. There is no fundamental difficulty in supporting existing software frameworks and inference-time optimizations like quantization with a minimal amount of engineering effort involved. Moreover, the increasing LLM inference demand can amortize these costs through massive production and deployment of these chips.